\documentclass[a4paper,12pt]{article}
\usepackage{epsfig}
\usepackage{citesort}
\usepackage{psfrag}
\date{\today}
 \normalsize

\def\be{\begin{equation}}
\def\ee{\end{equation}}
\def\bea{\begin{eqnarray}}
\def\eea{\end{eqnarray}}

\def\lsim{\raise0.3ex\hbox{$\;<$\kern-0.75em\raise-1.1ex\hbox{$\sim\;$}}}
\def\gsim{\raise0.3ex\hbox{$\;>$\kern-0.75em\raise-1.1ex\hbox{$\sim\;$}}}
\def\Frac#1#2{\frac{\displaystyle{#1}}{\displaystyle{#2}}}

\def\sm{\mbox{\tiny SM}}

\def\ddLR{(\delta^d_{LR})_{23}}

\def\duLLtwo{(\delta^u_{LL})_{32}}

\def\duLRtwo{(\delta^u_{LR})_{32}}

\begin{document}
\renewcommand{\thefootnote}{\fnsymbol{footnote}}
\rightline{hep-ph/0508024} \rightline{\today} \vspace{.3cm}
{\large
\begin{center}
{\bf Probing Flavor Structure in Supersymmetric Theories}
\end{center}}
\vspace{.3cm}
\begin{center}
Shaaban Khalil\\
\vspace{.3cm} \emph{Department of Mathematics, Ain Shams
University, Faculty of Science, Cairo, 11566, Egypt.}\\
\emph{Department of Mathematics, German University in Cairo, New
Cairo city, El Tagamoa El Khames, Egypt.}\\
\end{center}
\vspace{.3cm} \hrule \vskip 0.3cm
\begin{center}
\small{\bf Abstract}\\[3mm]

\begin{minipage}[h]{14.25cm}
We analyze the possibility of probing the supersymmetric flavor
structure through the constraints of the $K$ and $B$ meson systems
and those of the electric dipole moments. We show that combining
these constraints would favor SUSY models with large flavor mixing
either in $LR(RL)$ or $LL$ but with a very small $RR$ and
intermediate/large $\tan \beta$. Large $LR$ mixing requires
specific patterns for trilinear $A$-terms, while $LL$ mixing seems
quite natural and easier to obtain. We present an example for this
class of models and show how it can accommodate the current CP
asymmetries experimental results.
\end{minipage}
\end{center}
\vskip 0.3cm \hrule \vskip 0.5cm
\section{\large{\bf Introduction}}
Current data from $B$-factories on the branching ratios and the CP
asymmetries of $B\to \phi K$, $B\to \eta' K$ and $B\to K \pi$
suggest new sources of flavor and/or CP violation beyond the
Standard Model (SM). An attractive possibility for these new
sources can be found in supersymmetric (SUSY) models. These new
flavor and CP violation have significant implications and can
modify the SM predictions in flavor changing rare processes and CP
violating phenomena. However, experimental bounds on the electric
dipole moment (EDM) of the neutron, electron and mercury atom
usually impose stringent constraints on mixings and phases in the
adopted models. Therefore it is a challenge for SUSY models to
give a new source of flavor and CP that can explain the possible
discrepancy between CP asymmetry measurements and the expected SM
results, whilst at the same time avoiding the production of EDMs.

It is now clear that in order to accommodate the CP asymmetries of
different $B$ decays, SUSY models with flavor non-universal soft
breaking terms are favored. In this class of models, nontrivial
flavor structures in the squark mass matrices are obtained, and as
a result new flavor mixing and CP violation effects are expected
beyond those in the Yukawa couplings. However there is an open
debate about the type of the new flavor that one needs to
accommodate the current $B$ physics experimental results. The
squark mixings can be classified, according to the chiralities of
their quark superpartners, into left-handed or right-handed ($L$
or $R$) squark mixing. The $LL$ and $RR$ mixings represent the
chirality conserving transitions in the left- and right-handed
squarks and are given by the mass insertions
$(\delta^{u,d}_{LL})_{ij}$ and $(\delta^{u,d}_{RR})_{ij}$
respectively. The $LR$ and $RL$ refer to the chirality flipping
transitions and are given by the mass insertions
$(\delta^{u,d}_{LR})_{ij}$ and $(\delta^{u,d}_{RL})_{ij}$.

In the minimal flavor SUSY models, {\it i.e.}, SUSY models with
universal soft breaking terms, the $L$ and $R$ sectors of the up
and down squark matrices remain diagonal at the electroweak scale
to a very good approximation. Hence, this class of models can not
give any genuine contribution to the CP violating and flavor
changing processes in $K$ and $B$ systems \cite{Khalil:1999zn}.
The situation is drastically changed within the non-minimal flavor
SUSY models. Depending on the type of soft SUSY breaking, a large
mixing can be generated in these sectors. However, each sector is
severely constrained by flavor and/or CP violation experimental
limits . For instance, the mass insertions in the $LR$ and $RL$
are constrained by the EDMs, $\varepsilon'/\varepsilon$ and
$BR(b\to s \gamma)$ results, while the corresponding ones in the
$LL$ and $RR$ are constrained by $\Delta M_{K}$, $\Delta M_{B_d}$,
and $\varepsilon_k$ \cite{Gabbiani:1996hi}.

A salient feature of these constraints is that they are
generically more stringent on the $LR$ ($RL$) mass insertions than
the $LL$ ($RR$) mass insertions. Also, the transitions between
first and second generations in each sector are severely
constrained compared to those between first or second and third
generations. This gives the hope that SUSY contributions to the
$B$-system could be significant and may constitute an important
factor in explaining the current experimental results which show
some discrepancies from the SM predictions.

In this paper we pursue the discussion on the type of the SUSY
flavor which may contribute significantly to the CP asymmetries of
various $B$ decays without conflicting with the EDMs or any other
experimental results. We show that the scenario with large
$(\delta^d_{LR})_{23}$ and $(\delta^d_{RL})_{23}$ is consistent
and can give a solution to the CP asymmetry results. However, it
requires specific patterns for the non-universal trilinear
$A$-terms in order to avoid the stringent EDM constraint. One can
get another possible consistent solution through a large
$(\delta^d_{LL})_{23}$, but with a very small
$(\delta^d_{RR})_{23}$ and intermediate or large $\tan \beta$.
This type of models seems natural and can be obtained by a minimal
relaxation for the universality assumption of the minimal
supersymmetric standard model (MSSM). Moreover, large $\tan \beta$
is also favored by other experimental results like the branching
ratio of $B\to \mu^+ \mu^-$ \cite{Dedes:2003kp}.

The paper is organized as follows. In section 2 we make a critical
comparison between the two scenarios of large
$(\delta^d_{LR})_{23}$ and large $(\delta^d_{LL})_{23}$. In
section 3 we present an example for non-minimal flavor SUSY
models, where the scalar mass of the first two generations is
different from the scalar mass of the third generation. We also
show that this model can successfully pass the test of FCNC
constraints come from the kaon system. Section 5 is devoted to the
results of this model for the CP asymmetries of B-processes, in
particular the $B\to K \phi$, $B\to K \eta'$ and $B\to K \pi$. Our
conclusions are given in section 5.

\section{\large{\bf Squark mixing: LL versus LR mixing}}

It has been recently demonstrated that the EDM constraints
severely restrict the $LL$ and $RR$ contributions to the CP
asymmetries of $B\to \phi K$ and $B\to \eta' K$
\cite{Abel:2004te,Hisano:2004tf,Endo:2003te}. It was also pointed
out that SUSY models with dominant $LR$ and $RL$ mixing through
the non-universal $A$-terms may be the most favorite scenario to
accommodate the apparent deviation of the CP asymmetries from
those expected in the SM without contradicting the experimental
limits of EDMs \cite{Abel:2004te}. It is important to note that
these conclusions are based on the assumption of considering a
single mass insertion. The effect of large $(\delta^d_{LR})_{23}$
on the CP asymmetries of $B$ decays, particularly $B\to \phi K$,
$B\to \eta' K$ and $B\to K \pi$ has been systematically analyzed
\cite{susykphi,Gabrielli:2005ys,Khalil:2003bi,Khalil:2004yb,Khalil:2005qg}
and it was emphasized that it could naturally explain the observed
CP asymmetry results.

It is worth remembering that in the usual SUSY models, it is
rather difficult to arrange for a large mass insertion
$(\delta^d_{LR})_{23}\sim {\mathcal O}(10^{-2})$ whilst
maintaining the mass insertion $(\delta^d_{LR})_{12}$ small to
satisfy the constraints of $\Delta M_K$ and
$\varepsilon'/\varepsilon$: \be\rm{Re}(\delta^d_{LR})_{12} \lsim
{\mathcal O}(10^{-4}) ~~ \& ~~ \rm{Im}(\delta^d_{LR})_{12} \lsim
{\mathcal O}(10^{-5}).\ee Since the mass insertions
$(\delta^d_{LR})_{ij}$ are given by \bea (\delta^d_{LR})_{ij}
\simeq \left[V^{d^{\dagger}}_L . (Y^d A^d). V^d_R \right]_{ij}
~~~~~~ (\rm{for} ~ i\neq j) ,\label{deltaLR}\eea where $V^d_{L,R}$
are the diagonalization of the down quark mass matrix, all off
diagonal mass insertions would be, in principle, of the same order
unless one assumes a very specific flavor structure for the
$A$-terms. In fact the factorizable $A$-term that has been
considered in Ref.\cite{Khalil:2000ci,Khalil:2003mz} is an example
of this type of pattern that may lead to such a hierarchy between
$(\delta^d_{LR})_{23}$ and $(\delta^d_{LR})_{12}$. Moreover, one
needs to assume non-hierarchical Yukawa textures to avoid a
possible suppression for the off-diagonal entries of the mass
insertions which, as can be seen from Eq.(\ref{deltaLR}), depend
on the corresponding Yukawa couplings. Therefore, it is not an
easy task to get $(\delta^d_{LR})_{23}$ of order $10^{-2}$.

However, it was realized that with intermediate/large $\tan
\beta$, the double mass insertions could be quite relevant and may
lead to an effective $(\delta^d_{LR})_{23}$ of the required order
even with universal $A$-terms \cite{Endo:2004dc}. This can be seen
from the explicit dependence of $(\delta_{LR(RL)}^d)_{23}$ on the
$LL(RR)$ mixing, which is give by \bea
(\delta^d_{LR})_{23_{\rm{eff}}} = (\delta^d_{LR})_{23} +
(\delta^d_{LL})_{23}~ (\delta^d_{LR})_{33}, \eea where
$(\delta^d_{LR})_{33}\simeq \frac{m_b (A_b - \mu \tan
\beta)}{\tilde{m}^2}$. Thus if the mass insertion
$(\delta^d_{LR})_{23}$ is negligible one finds \bea
(\delta^d_{LR})_{23_{\rm{eff}}} \simeq (\delta^d_{LL})_{23}~
\frac{m_b}{\tilde{m}}~ \tan\beta.\eea Here we assumed that $\mu
\sim \tilde{m}$ and the phase of $\mu$ set to zero to overcome the
EDM constraints. It is clear that with $(\delta^d_{LL})_{23}
\simeq 10^{-2}$ one can easily get
$(\delta^d_{LR})_{23_{\rm{eff}}}$ of order $10^{-3} - 10^{-2}$,
depending on the value of $\tan \beta$. Similarly, one can
generate an effective $(\delta^d_{RL})_{23}$ of the right order
through large $(\delta^d_{RR})_{23}$.

In Ref. \cite{Endo:2004dc}, this contribution has been considered
as an $LL$ contribution to the CP asymmetry of $B$ decay. This
identification was given to indicate the type of large mixing in
the squark mass matrix. Nevertheless we should be aware that the
main effect of SUSY contribution is still due to the Wilson
coefficient $C_{8g}$ of the chromomagnetic operator, which is
enhanced by the chirality flipped factor $m_{\tilde{g}}/m_b$. It
is also worth mentioning that it is quite natural in SUSY models
to achieve $LL$ mixing between the second and third families of
order $10^{-2}$. Although this size of mixing is not enough to
explain the measured values of the CP asymmetries of $B$-decays,
yet it could induce an effective $LR$ mixing that accounts for
these results.

Having said that though, one should be very careful with the EDM
constraints. The mass insertion $(\delta^d_{LR})_{22}$, which is
severely constrained by the experimental limit on the mercury EDM
\cite{Abel:2001vy}:
$$\rm{Im}(\delta^d_{LR})_{22} < 5.6 \times 10^{-6}$$ can be
overproduced and thus may violate this bound. As explained in
Ref.\cite{Abel:2004te}, the effective mass insertion
$(\delta^d_{LR})_{22_{\rm{eff}}}$ can be expressed as \bea
(\delta^d_{LR})_{22{\rm{eff}}} &\simeq& 10^{-2}~ \tan \beta
\left[(\delta^d_{LL})_{23}~ (\delta^d_{RR})^*_{23} +
\left((\delta^d_{RR})_{23}~ (\delta^d_{LL})^*_{23} \right)^*
\right]. \eea  Hence, in this scenario it is necessary to have
either $(\delta^d_{LL})_{23}$ or $(\delta^d_{RR})_{23}$ less than
$10^{-3}$. For instance with $\tan \beta=10$, one should have
$(\delta^d_{LL(RR)})_{23} \simeq {\cal O} (10^{-1})$ so that
$(\delta^d_{LR})_{23_{\rm{eff}}}\simeq {\cal O}(10^{-2})$ to
accommodate the CP asymmetries and $(\delta^d_{RR(LL)})_{23} <
10^{-4}$ to avoid the mercury EDM constraint. It is known that in
MSSM with universal boundary condition, the mass insertion
$(\delta^d_{LL})_{23}$ is of order $10^{-3}$. This value can be
considered as a lower limit to the $(\delta^d_{LL})_{23}$,
therefore it is clear that models with large $RR$ mixing would be
disfavored by the EDM constraints
\cite{Abel:2004te,Hisano:2004tf,Endo:2003te}.

Another argument which also motivates the class of SUSY models
with large $LL$ mixing is the fact that both this mixing and the
intermediate/large values of $\tan \beta$ are essential
requirements for enhancing the chargino contributions which play a
crucial role in explaining the experimental results of $B\to K
\pi$ branching ratio and CP asymmetries
\cite{Khalil:2004yb,Khalil:2005qg}. Note that due to the $SU(2)$
gauge invariance the soft scalar masses $M_Q^2$ is the same for
the up and down sectors. Hence, the up and down mass insertions
are related as follows: \bea
 (\delta^d_{LL})_{ij} = \left[ V_{CKM}^+
(\delta^u_{LL}) V_{CKM}\right]_{ij}, \eea
 {\it i.e.},
\bea (\delta^d_{LL})_{23} = (\delta^u_{LL})_{23} + \lambda
(\delta^u_{LL})_{13}+ {\mathcal O}(\lambda^2), \eea with $\lambda
=0.22$. Therefore, a non-universal $M_Q^2$ can lead to large
$(\delta^d_{LL})_{23}$ and $(\delta^u_{LL})_{23}$. In this
respect, this scenario is very economical in that it can explain
many results with quite few assumptions.

\section{\large{\bf Suggested supersymmetric flavor model}}
As advocated above, the non-universal soft breaking terms are
crucial ingredients to have a new flavor structure beyond the
usual Yukawa couplings and to enhance the effect of the SM phase
$\delta_{CKM}$. Moreover, general supergravity models and most of
string and $D$-brane inspired models naturally lead to
non-universal soft SUSY breaking parameters
\cite{strngnonuniversality}. The soft scalar masses of the first
two generations are generally assumed degenerate in order to avoid
the flavor changing neutral current (FCNC) constraints, especially
the $\Delta M_K$ and $\varepsilon_K$ which impose very strong
constraints on (12) mixings. As an example, we consider here a
SUSY model with the following soft breaking terms at the GUT scale
\bea && M_1=M_2 =M_3 = M_{1/2}
~~~~~~~~~~~~~~ (\rm{universal~ gaugino~ mass}),\\
&& A^u = A^d = A_0 ~~~~~~~~~~~~~~~~~~~~~~~~~~ (\rm{universal~ A-term}),\\
&&M_{U}^2 = M_{D}^2 = m_0^2 ~~~~~~~~~~~~~~~~~~~~~~~
(\rm{universal~ mass~
for~ the~ squark~ singlets}),\\
&& m_{H_1}^2 = m_{H_2}^2 = m_0^2 ~~~~~~~~~~~~~~~~~~~~~
(\rm{universal~ Higgs~ masses}). \eea The masses of the squark
doublets are given by
\begin{equation}
M_{Q}^2 = \left( \matrix{ m_0^2 &  & \cr
 & m_0^2 &  \cr
 &  & a^2 m_0^2 }\right) \;.
 \label{MQ}
\end{equation}
The parameter $a$ measures the deviation between the masses of the
third and the first two generations. This model is a special case
of texture (C) that has been recently studied in
Ref.\cite{Chankowski:2005jh}.

Given the above boundary condition for the soft terms at the GUT
scale, we determine the evolution of the various couplings
according to their renormalization group equations. At the weak
scale, we impose the electroweak symmetry breaking conditions and
calculate the Higgsino mass $\mu$ (up to a sign) and the bilinear
parameter $B$. This imposes a constraint on the parameter $a$. We
will assume through the paper the following values: $\tan\beta=15$
and $m_0=M_{1/2}=A_0=250$ GeV. For these values $a$ has an upper
bound $ a \leq 5$. The sparticle spectrum is explicitly computed
at the weak scale in terms of the parameters: $M_{1/2},m_0,A_0,a$,
and $\tan\beta$. With non-universal soft SUSY breaking terms, the
Yukawa textures play an important rule in the CP and flavour
supersymmetric results and one has to specify the type of the
Yukawa couplings in order to completely determine the model. Here
we will use the following simple Yukawa textures given in terms of
the quark masses and CKM mixing matrix: \bea
Y^u &=& \frac{1}{v\sin \beta} ~ \rm{diag}\left(m_u,m_c,m_t\right),\\
Y^d &=&  \frac{1}{v\sin \beta} ~ V^{\dagger}_{CKM}\cdot\rm{diag}
\left(m_d ,m_s,m_b\right) \cdot V_{CKM}. \eea This type of Yukawa
texture is hierarchical, so it is not the best choice since it
dilutes the effect of the SUSY flavor. However, as we will show,
this texture gives good results for flavor mixing between the
second and third generation in the squark mass matrices.

Although, a very useful tool for analyzing SUSY contributions to
FCNC processes is provided by the mass insertion approximation,
one should be careful in models with non-universal soft terms. In
our model, with $a\neq 1$, we get a highly non-degenerate spectrum
which violates one of the assumptions of the mass insertion
approximation. Therefore, in our analysis we will use the full
loop computation. Nevertheless, it may be still useful to consider
the mass insertions just to understand the main features of this
model and how it differs from the other models with non-universal
$A$-terms. The $LL$ down mass insertions are defined in the
super-CKM basis, as \bea(\delta^{d}_{LL})_{ij} =
\frac{1}{\tilde{m}^2} \left[ V_L^{d^{\dagger}} ({\cal M}^d)^2_{LL}
V^{d}_L\right]_{ij},\eea where $({\cal M}^d)^2_{LL}$ is the $LL$
down squark at the electroweak scale, $\tilde{m}$ is the average
of the squark mass, and $V^d_L$ is the rotation matrix that
diagonalizes the down quark mass matrix. Thus, for the soft scalar
masses $M_Q^2$ given in Eq.(\ref{MQ}) and $a=5$, one finds \bea
(\delta^d_{LL})_{23} \simeq 0.08 ~ e^{0.4 i}.
\label{herresult}\eea Although we are using a hierarchical Yukawa
texture, the result looks very promising. It is clear that with
such value of $(\delta^d_{LL})_{23}$, one can easily get
$(\delta^d_{LR})_{23_{\rm{eff}}}\simeq {\cal O}(10^{-2}-
10^{-3})$. Recall that the corresponding single $LR$ mass
insertion is negligible due to the degeneracy of the $A$-terms.
Finally, we also find that the $(\delta^d_{LL})_{12}$ is given by
 \be
(\delta^d_{LL})_{12} \simeq 0.0002 + 0.0002 i .\ee This result
satisfies the strongest constraints coming from the kaon physics:
$\sqrt{\vert \rm{Re}(\delta^d_{LL})_{12}^2\vert} \lsim 4\times
10^{-2}$ which is imposed by the measured value of $\Delta M_K$
and $\sqrt{\vert \rm{Im}(\delta^d_{LL})_{12}^2\vert} \lsim 4\times
10^{-3}$ from $\varepsilon_K$. Since $(\delta^d_{LR})_{22} \simeq
4 \times 10^{-3}$, the imaginary part of the effective mass
insertion $(\delta^d_{LR})_{12_{\rm{eff}}}$ is given by \be
\rm{Im}\left[(\delta^d_{LR})_{12_{\rm{eff}}}\right] \simeq
10^{-6}, \label{eps'} \ee which satisfies the bound imposed by
$\varepsilon'/\varepsilon$: $\vert~
\rm{Im}(\delta^d_{LR})_{12}\vert \lsim 2\times 10^{-5}$. Note that
in this case both of $\Delta M_K$, $\varepsilon_K$ and
$\varepsilon'/\varepsilon$ should be saturated by the SM
contribution. However, it is quite possible to enhance the SUSY
contribution, if necessary, by considering more non-hierarchial
Yukawa texture.

Before we proceed and determine the SUSY contributions to the CP
asymmetries of $B$ processes, one important remark is in order.
This model, like the constrained MSSM, has in general two
independent phases: $\phi_A$ and $\phi_{\mu}$. However, these two
phases are strongly constrained by the EDM. Therefore, we set them
to zero and assume that the SUSY breaking mechanism is preserving
the CP violation. Hence, the only source of CP violation here is
the SM phases $\delta_{CKM}$. In the sprit of
Ref.\cite{Khalil:2003mz}, we will show that the new source of SUSY
flavor with $\delta_{CKM}$ is sufficient to accommodate the
current experimental results.

\section{\large{\bf Contribution to the CP asymmetry of $B$ processes}}
The most recent results of BaBar and Belle collaborations
\cite{Abe:2005bt,Aubert:2005iy} on the mixing-induced asymmetries
of $B\to \phi K$ and $B\to \eta' K$ are given as follows: The
Belle experimental values of these asymmetries are \bea S_{\phi K}
&=&
0.44\pm 0.27\pm 0.05, \\
S_{\eta' K} &=& 0.62\pm 0.12 \pm 0.04. \label{belleresult}\eea The
BaBar experimental results are \bea S_{\phi K} &=& 0.50\pm
0.25^{+0.07}_{-0.04}, \\
S_{\eta' K} &=& 0.30\pm 0.14 \pm 0.02. \label{babarresult}\eea
Comparison with the world average CP asymmetry of $B\to J/\psi K$,
which is now given by $S_{J/\psi K}=0.685\pm0.032$, shows that the
average $S_{\phi K_S} = 0.47 \pm 0.19.$ displays about $1 \sigma$
deviation from SM prediction, while the average $S_{\eta' K_S} =
0.48 \pm 0.09$ displays $2.5 \sigma$ discrepancy.

On the other hand the latest experimental results for the direct
CP violation in $\bar{B}^0 \to K^- \pi^+$ and $B^- \to K^- \pi^0$
are given by \cite{Abe:2005fz}
\bea
A^{CP}_{K^- \pi^+} &=& -0.113 \pm 0.019 \\
A^{CP}_{K^- \pi^0} &=&  0.04 \pm 0.04. \eea The result of
$A^{CP}_{K^- \pi^+}$ corresponds to a $4.2 \sigma$ deviation from
zero, while the measured value of $A^{CP}_{K^- \pi^0}$, which may
also exhibit a large asymmetry, is quite small. These observations
have has been considered as possible signals to new physics
\cite{Khalil:2005qg,Buras:2004th}. In this section we will study
the contribution of our SUSY model to these CP violating
asymmetries.

\subsection{\large{\bf Contributions to $S_{\phi K}$ and $S_{\eta'
K}$}}

As can be seen from Eqs.(\ref{belleresult}-\ref{babarresult}), it
seems that the CP asymmetry $S_{\phi K}$ is consistent with the SM
result and SUSY contributions should be within the experimental
errors. The situation of $S_{\eta' K}$ is not yet clear for the
following two reasons. First Bell and BaBar still give quite
different results. Second, it is commonly believed that $\eta'$ is
a more complicated particle than $\phi$ and its CP asymmetry could
be different due to some peculiar dynamics for this particle. In
any case, we will consider here $S_{\phi K}$ as a constraint and
will study the possible prediction of our SUSY models for
$S_{\eta' K}$ and also for the direct CP asymmetries of $B\to K
\pi$ decays.

As emphasized in Refs.\cite{Gabrielli:2005ys}, the dominant gluino
contributions are due to the QCD penguin diagrams and
chromo-magnetic dipole operators. The gluino contributions to the
corresponding Wilson coefficients at the SUSY scale can be found
in Ref.\cite{Harnik:2002vs}. The $LR$ contributions only enter the
Wilson coefficients $C_{7\gamma}$ and $C_{8g}$ of the magnetic and
chromo-magnetic operators:
\begin{eqnarray}
C^{\tilde{g}}_{7\gamma}  &=& \frac{\alpha_s \pi}{m_{\tilde{g}}^2}
\left[ \sum_{AB} \Gamma_{sA}^{R^*} \Gamma_{bA}^{R}\left(
\frac{-4}{9} D_1(x_A) \right) + \frac{m_{\tilde{g}}}{m_b} \sum_A
\Gamma_{sA}^{R^*} \Gamma_{sA}^{L} \left(-\frac{4}{9} D_2(x_A)\right)\right],\\
C^{\tilde{g}}_{8g} \!\! &=&\!\!\frac{\alpha_s
\pi}{m_{\tilde{g}}^2}\!\left[\sum_{AB}\!\Gamma_{sA}^{R^*}
\Gamma_{bA}^{R}\!\left(\!\frac{-1}{6}D_1(x_A)\!+\!\frac{3}{2}
D_3(x_A)\!\right)\!\!+\!\frac{m_{\tilde{g}}}{m_b}\sum_A
\Gamma_{sA}^{R^*}\!\Gamma_{sA}^{L}\!\left(\!-\frac{1}{6}
D_2(x_A)\!+\!\frac{3}{2}D_4(x_A)\!\right)\!\right]\!\nonumber,
\label{Cgluino}
\end{eqnarray}
where $x_A =\tilde{m}_A^2/m_{\tilde{g}}^2$ and the loop functions
are given in Ref.\cite{Harnik:2002vs}. In our numerical analysis,
we include Wilson coefficients of all the relevant operators and
the ones obtained from these operators by the chirality exchange.
In our discussion we will focus on $C^{\tilde{g}}_{7\gamma}$ and
$C^{\tilde{g}}_{8g}$ which give the dominant contribution due to
the large enhancement factor $m_{\tilde{g}}/m_b$ in front of the
term proportional to the LR mixing.

We will apply the QCD factorization which allows to estimate the
hadronic matrix elements of the involved operators. In this case,
the SUSY contribution to the decay amplitude of $B\to \phi K$ is
given by \cite{Gabrielli:2005ys}
\begin{equation}
A(B\to \phi K) \simeq - i \frac{G_F}{\sqrt{2}} m_B^2 F_+^{B\to K}
f_{\phi} H_{8g} \left(C_{8g} + \tilde{C}_{8g}\right). \label{APhi}
\end{equation}
Here $m_{\phi}=1.02$ GeV is the $\phi$ meson mass, $F_+^{B\to
K}=0.35\pm 0.05$ is the transition form factor evaluated at
transferred momentum of order $m_{\phi}$, and $f_{\phi}=0.233$ GeV
is the $\phi$ meson form factor. The coefficient $H_{8g}$ is given
by $H_{8g} = 0.047$ \cite{Gabrielli:2005ys}. Note that
$H_{7\gamma}$ is two order of magnitude smaller than $H_{8g}$,
therefore we neglect the magnetic moment dipole contribution.
Since the hard scattering and weak annihilation contributions to
$Q_{8g}$ have not been calculated, the coefficient $H_{8g}$ has no
strong phase dependence. It is expected that this contribution has
an undetermined strong phase. This will increase the theoretical
uncertainty since $Q_{8g}$ is giving the dominant contribution in
SUSY model. Here, we assume that the matrix element of $Q_{8g}$
induces a strong phase $\delta_{\phi}$ to the SUSY contribution to
the $B\to \phi K$ amplitude. Thus, the ratio of the SUSY and SM
amplitudes can be written as \be
\left(\frac{A^{\rm{SUSY}}}{A^{\rm{SM}}}\right)_{\phi K} = R_{\phi}
~e^{i \theta_{\phi}} ~e^{i\delta_{\phi}},\ee where $R_{\phi}$
stands for $\left\vert
\left(\frac{A^{\rm{SUSY}}}{A^{\rm{SM}}}\right)_{\phi K}
\right\vert$ and $\theta_{\phi}$ for the $\rm{Arg}[C_{8g}]$ since
$\tilde{C}_{8g}$ is negligible with respect to $C_{8g}$ in this
class of model. Similarly, the SUSY contribution to the decay
amplitude of $B\to \eta' K$ is given by \cite{Gabrielli:2005ys}
\begin{equation}
A(B\to \eta' K) \simeq - i \frac{G_F}{\sqrt{2}} m_B^2 F_+^{B\to K}
f_{\eta'} H'_{8g} \left(C_{8g} - \tilde{C}_{8g}\right),
\label{Aeta}
\end{equation}
and the ratio of the SUSY and SM amplitudes can be written as \be
\left(\frac{A^{\rm{SUSY}}}{A^{\rm{SM}}}\right)_{\eta' K} =
R_{\eta'} ~e^{i \theta_{\eta'}} ~e^{i\delta_{\eta'}},\ee where
$R_{\eta'}$ refers to $\left\vert
\left(\frac{A^{\rm{SUSY}}}{A^{\rm{SM}}}\right)_{\eta'K}
\right\vert$, $\theta_{\eta'}\simeq \theta_{\phi}\simeq
\rm{Arg}[C_{8g}]$, and $H'_{8g}=-0.89$.

\begin{figure}[t]
\begin{center}
\includegraphics[width=10.5cm]{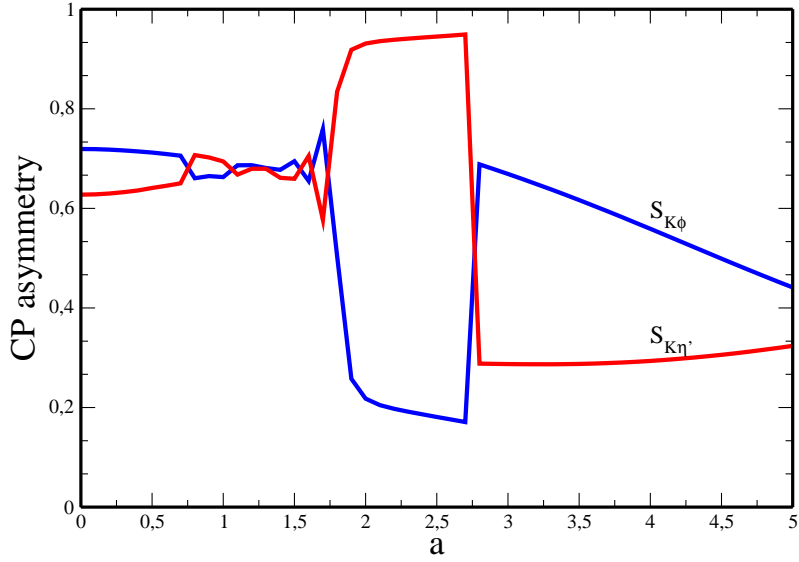}
\end{center}
\caption{\small CP asymmetries of $B\to \phi K$ and $B\to \eta' K$
as function of the squark non-universality parameter $a$ for $\tan
\beta =15$, $m_0=M_{1/2}=A_0=250$ GeV, $\delta_{\phi}\sim 2/3
\pi$, and $\delta_{\eta'}=0$.} \label{Fig1}
\end{figure}
As a result of small $RR$ mixing in the class of models that we
consider, the sign difference between $C_i$ and $\tilde{C}_i$ in
$B\to \eta' K$ transition \cite{Khalil:2003bi}, can not be used to
create a significant difference between $A^{\rm{SUSY}}_{\phi K}$
and $A^{\rm{SUSY}}_{\eta' K}$. However, as we will show, due to
the fact that the strong phases in $B\to \phi K$ and $B\to \eta'
K$ are in general different, one can get the required deviation
between $S_{\phi k}$ and $S_{\eta' K}$. Using the above
parametrization of the SM and SUSY amplitudes, the mixing CP
asymmetries $S_{\phi (\eta')K}$ can be written as \be S_{\phi
(\eta') K}=\Frac{\sin 2 \beta +2 R_{\phi (\eta')} \cos
\delta_{\phi(\eta')} \sin(\theta_{\phi (\eta')} + 2 \beta) +
R_{\phi (\eta')}^2 \sin (2 \theta_{\phi (\eta')} + 2 \beta)}{1+ 2
R_{\phi (\eta')} \cos \delta_{\phi(\eta')} \cos\theta_{\phi
(\eta')} +R_{\phi (\eta')}^2}~. \label{CPasym} \ee

In Fig.1 we present the CP asymmetries $S_{\phi K}$ and $S_{\eta'
K}$ as function of the non-universality parameter $a$ for
$m_0=M_{1/2}=A_0= 250$ GeV and $\tan \beta=15$. Also the strong
phases are fixed as $\delta_{\phi}\simeq 2\pi/3$ while
$\delta_{\eta'} = 0$. As can be seen from this figure, in this
class of models with a large $a$, it is possible to account
simultaneously for the experimental results of $S_{K\phi}$ and
$S_{K\eta'}$.

\subsection{\large{\bf Contributions to $B\to K \pi$}}
Now let us turn to the gluino contribution to $B\to K^-\pi^+$ and
$B\to K^-\pi^0$. As emphasized in Ref.\cite{Khalil:2005qg}, the
direct CP asymmetries of $B\to K \pi$ decays can be approximately
given by \bea A^{CP}_{K^-\pi^+} &\simeq & 2 r_T \sin \delta_T
\sin(\theta_P + \gamma) +2 r_{EW}^C \sin \delta_{EW}^C
\sin(\theta_P -
\theta_{EW}^c),\label{CP1}\\
A^{CP}_{K^-\pi^0} &\simeq & 2 r_{T} \sin \delta_{T}
\sin(\theta_P+\gamma) - 2 r_{EW}\sin \delta_{EW} \sin(\theta_P-
\theta_{EW}).\label{CP2} \eea The parameters $\theta_P$,
$\theta_{EW}$, $\theta_{EW}^c$ and $\delta_T$, $\delta_{EW}$,
$\delta^c_{EW}$ are the CP violating and CP conserving (strong)
phases respectively. The parameters $r_T$ measures the relative
size of the tree and QCD penguin contributions. While $r_{EW},
r_{EW}^C$ measure the relative size of the electroweak and QCD
contributions. By assuming the same strong phases for SM and SUSY
contribution, we can write \cite{Khalil:2004yb,Khalil:2005qg} \bea
P e^{i\theta_P}&=& P^{\sm} (1+ke^{i\theta^{\prime}_P}), \\
r_{EW} e^{i\delta_{EW}}e^{i\theta_{EW}}&=&
(r_{EW})^{\sm}e^{i\delta_{EW}}
(1+le^{i\theta^{\prime}_{EW}}), \label{rEW}\\
r_{EW}^C e^{i\delta_{EW}^C}e^{i\theta_{EW}^C}&=&
(r_{EW}^{C})^{\sm}e^{i\delta_{EW}^C}(1+me^{i\theta^{C^\prime_{EW}}}),\label{rEWC}\\
r_T e^{i\delta_T}&=&\frac{(r_T e^{i\delta_T})_{\sm}}{\left|1+k
e^{i\theta^{\prime}_P}\right|} \label{risusy} \eea where $k,l,m$
are given in terms of the $(\delta^d_{LR})_{23}$ through gluino
contributions and $(\delta^u_{LL})_{32}$ and
$(\delta^u_{LR})_{32}$ through chargino contributions. For glunio
mass of order $500$ GeV, $m_{\tilde{q}} =500$ GeV,
$m_{\tilde{t}_R}=150$ GeV, $M_2=200$ GeV and $\mu =400$ GeV, one
finds \cite{Khalil:2004yb,Khalil:2005qg} \bea
ke^{i\theta_P}&=& -0.0019 \tan\beta \duLLtwo - 35.0 \ddLR +0.061\duLRtwo\label{eq:kSUSY}\\
le^{i\theta_q}&=& 0.0528 \tan\beta \duLLtwo-2.78\ddLR +1.11\duLRtwo \label{eq:lSUSY}\\
me^{i\theta_{q_C}}&=& 0.134 \tan\beta\duLLtwo + 26.4\ddLR +1.62
\duLRtwo. \label{eq:mSUSY} \eea Since we have assumed a diagonal
up-Yukawa couplings, the flavor mixing among the up squarks is
very small. The typical values of the mass insertion
$(\delta^u_{LL})_{32}$ and $(\delta^u_{RL})_{32}$ are of order
$10^{-3}$ , so that the chargino contribution is negligible. On
the other hand with $a=5$ and $m_0=M_{1/2}=A_0=250$ GeV, the mass
insertion $(\delta^d_{LR})_{32}$ is give by
$(\delta^d_{LR})_{32}\simeq 0.006 \times e^{-2.7~i}$. Therefore
one finds \be k\simeq0.2 ~~~~~~~ l\simeq0.009 ~~~~~~~ m\simeq0.16
\ee From Eqs.(\ref{rEW}-\ref{risusy}), it is clear that in this
example, $r_{EW}$ and $r^c_{EW}$ are given, to a good
approximation, by the SM values: $r_{EW}^{SM} \simeq 0.13$ and
$(r^c_{EW})^{SM} \simeq 0.012$, while $r_T$ is reduced from
$r_T^{SM}\simeq 0.2$ to $r_T \simeq 0.16$. As explained in
Ref.\cite{Khalil:2005qg}, in this case with $r_T,r_{EW} \gg
r_{EW}^c$, the CP asymmetry $A^{CP}_{K^- \pi^+}$ is given by the
first term in Eq.(\ref{CP1}) which can easily be of order
$-0.113$. However, the CP asymmetry $A^{CP}_{K^- \pi^0}$ receives
contributions from both terms of Eq.(\ref{CP2}). With $r_T \sim
r_{EW}$, the possibility of having cancellation between these two
terms is quite large and one obtains $A^{CP}_{K^- \pi^0}<
A^{CP}_{K^- \pi^+}$, as required by the current experimental
results.

\section{\large{\bf conclusions}}

In this paper we have studied the possibility of probing the
supersymmetric flavor structure. We have used the experimental
constraints from the CP asymmetries of $K$ and $B$ meson systems
and also from the electric dipole moments. We have shown that
these constraints would lead together to a specific SUSY flavor
structure. One possibility is to have a large flavor mixing in
$LR$ and/or $RL$ sector. The second possibility is to have a large
mixing in $LL$ combined with a very small mixing in the $RR$
sector and also intermediate or large $\tan \beta$. The scenario
of large $LR$ mixing requires a specific pattern for trilinear
$A$-terms, like, factorizable or Hermitian $A$ terms for instance.
On the other hand $LL$ mixing scenario seems quite natural and can
be obtained by a non-universality between the squark masses. As an
example, we considered a SUSY model with a minimal relaxation for
the universality assumption of the MSSM, where the masses of the
left squarks of the first two generations and the third generation
are different. We have shown that in this class of models, one can
get effective mass insertion $(\delta_{LR}^d)_{23}$  that leads to
a significant SUSY contribution to the CP asymmetry of $B$ decays.
In particular, we have emphasized that the new results of $S_{\phi
k}$ and $S_{\eta' K}$ can be accommodated. Also the model can
account for the observed correlation between $A^{CP}_{K^-\pi^+}$
and $A^{CP}_{K^-\pi^0}$.



\begin{thebibliography}{99}
%
\bibitem{Khalil:1999zn}
  S.~Khalil, T.~Kobayashi and A.~Masiero,
  Phys.\ Rev.\ D {\bf 60}, 075003 (1999).
  D.~A.~Demir, A.~Masiero and O.~Vives,
  Phys.\ Lett.\ B {\bf 479}, 230 (2000).

\bibitem{Gabbiani:1996hi}
F.~Gabbiani, E.~Gabrielli, A.~Masiero and L.~Silvestrini,
Nucl.\ Phys.\ B {\bf 477}, 321 (1996).

\bibitem{Dedes:2003kp}
  A.~Dedes,
  Mod.\ Phys.\ Lett.\ A {\bf 18}, 2627 (2003), references therein.

\bibitem{Abel:2004te}
S.~Abel and S.~Khalil,
Phys.\ Lett.\ B {\bf 618}, 201 (2005).


\bibitem{Hisano:2004tf}
J.~Hisano and Y.~Shimizu,
Phys.\ Rev.\ D {\bf 70}, 093001 (2004);Phys.\ Lett.\ B {\bf 581},
224 (2004).

\bibitem{Endo:2003te}
M.~Endo, M.~Kakizaki and M.~Yamaguchi,
Phys.\ Lett.\ B {\bf 583}, 186 (2004).

\bibitem{susykphi}
E. Lunghi and D. Wyler, Phys. Lett. B 521 (2001) 320; M. B.
Causse, arXiv:hepph/ 0207070; G. Hiller, Phys. Rev. D 66 (2002)
071502; M. Ciuchini and L. Silvestrini, Phys. Rev. Lett. 89 (2002)
231802; S. Khalil and E. Kou, Phys. Rev. D 67 (2003) 055009; K.
Agashe and C. D. Carone, Phys. Rev. D 68 (2003) 035017; G. L.
Kane, P. Ko, H. b. Wang, C. Kolda, J. h. Park and L. T. Wang;
Phys. Rev. Lett. 90 (2003) 141803; C. Dariescu, M.A. Dariescu,
N.G. Deshpande, D.K. Ghosh, Phys. Rev. D 69 (2004) 112003; M.
Ciuchini, E. Franco, G. Martinelli, A. Masiero, M. Pierini, L.
Silvestrini, hep-ph/0407073. Z. Xiao and W. Zou, hep-ph/0407205;
D. Chakraverty, E. Gabrielli, K. Huitu and S. Khalil, Phys. Rev. D
68 (2003) 095004; S. Khalil and R. Mohapatra, Nucl. Phys. B 695
(2004) 313.

\bibitem{Gabrielli:2005ys}
E.~Gabrielli, K.~Huitu and S.~Khalil, Nucl.\ Phys.\ B {\bf 710},
139 (2005).

\bibitem{Khalil:2003bi}
S. Khalil and E. Kou, Phys. Rev. Lett. 91 (2003) 241602.

\bibitem{Khalil:2004yb}
S.~Khalil and E.~Kou,
Phys.\ Rev.\ D {\bf 71}, 114016 (2005).

\bibitem{Khalil:2005qg}
S.~Khalil,
arXiv:hep-ph/0505151.

\bibitem{Khalil:2000ci}
S.~Khalil, T.~Kobayashi and O.~Vives,
Nucl.\ Phys.\ B {\bf 580}, 275 (2000). T.~Kobayashi and O.~Vives,
Phys.\ Lett.\ B {\bf 506}, 323 (2001). D.~Bailin and S.~Khalil,
Phys.\ Rev.\ Lett.\  {\bf 86}, 4227 (2001).

\bibitem{Khalil:2003mz}
S.~Khalil and V.~Sanz,
Phys.\ Lett.\ B {\bf 576}, 107 (2003).

\bibitem{Endo:2004dc}
M.~Endo, S.~Mishima and M.~Yamaguchi,
Phys.\ Lett.\ B {\bf 609}, 95 (2005).

\bibitem{Abel:2001vy}
  S.~Abel, S.~Khalil and O.~Lebedev,
  Nucl.\ Phys.\ B {\bf 606}, 151 (2001).

\bibitem{strngnonuniversality}
A. Brignole, L. E. Ibanez , and C. Munoz , Nucl. Phys. B 422 (1994) 125,
Erratum-ibid. B 436 (1995) 747.
L. E. Ibanez , C. Munoz , and S. Rigolin, Nucl. Phys. B 553 (1999) 43.


\bibitem{Chankowski:2005jh}
  P.~H.~Chankowski, O.~Lebedev and S.~Pokorski,
  Nucl.\ Phys.\ B {\bf 717}, 190 (2005) .

\bibitem{Abe:2005bt}
  K.~Abe  [Belle Collaboration],
  arXiv:hep-ex/0507037.

\bibitem{Aubert:2005iy}
  B.~Aubert {\it et al.}  [BaBar Collaboration],
  Phys.\ Rev.\ Lett.\  {\bf 94}, 191802 (2005);
M. A.Giorgi (BaBar collaboration), plenary talk at XXXII Int.
Conference on High Energy Physics, Beijing, China, August 16-22,
2004, http://ichep04.ihep.ac.cn/

  \bibitem{Abe:2005fz}
  K.~Abe,
  arXiv:hep-ex/0507045.

\bibitem{Buras:2004th}
A.~J.~Buras, R.~Fleischer, S.~Recksiegel and F.~Schwab,
  Acta Phys.\ Polon.\ B {\bf 36}, 2015 (2005);
X.~G.~He and B.~H.~J.~McKellar,
  arXiv:hep-ph/0410098;
  A.~J.~Buras, R.~Fleischer, S.~Recksiegel and F.~Schwab,
  Nucl.\ Phys.\ B {\bf 697}, 133 (2004);
X. G. He and B. H. J. McKellar, arXiv:hep-ph/0410098; X. G. He, C.
S. Li and L. L. Yang, Phys. Rev. D 71, 054006 (2005); S. Mishima
and T. Yoshikawa, Phys. Rev. D 70, 094024 (2004); S. Baek, P.
Hamel, D. London, A. Datta and D. A. Suprun, Phys. Rev. D 71,
057502 (2005); A.J.Buras and R.Fleischer, Eur. Phys. J. C 16
(2000) 97; M.Gronau and J.L.Rosner, Phys. Lett. B 572 (2003) 43;
T. Yoshikawa, Phys. Rev. D 68 (2003) 054023; S. Nandi and A.
Kundu, arXiv:hep-ph/0407061; A. J. Buras, R. Fleischer, S.
Recksiegel and F. Schwab, Phys. Rev. Lett. 92 (2004) 101804 ;
arXiv:hep-ph/0402112; Y. Grossman, M. Neubert and A. L. Kagan,
JHEP 9910 (1999) 029; V. Barger, C. W. Chiang, P. Langacker and H.
S. Lee, arXiv:hep-ph/0406126; M. Ciuchini, E. Franco, G.
Martinelli, A. Masiero, M. Pierini and L. Silvestrini,
arXiv:hep-ph/0407073; W. S. Hou, M. Nagashima and A. Soddu,
arXiv:hep-ph/0503072.

\bibitem{Harnik:2002vs}
R.~Harnik, D.~T.~Larson, H.~Murayama and A.~Pierce,
Phys.\ Rev.\ D {\bf 69}, 094024 (2004).

\end{thebibliography}
\end{document}